\documentclass{article}
\usepackage{arxiv}
\usepackage{natbib}
\usepackage{graphicx}
\usepackage{epstopdf, epsfig}
\usepackage{amssymb}
\usepackage{amsmath}
\usepackage{hyperref}
\usepackage[nameinlink, noabbrev]{cleveref}    
\usepackage{xcolor}
\usepackage{subfig}
\usepackage{float}
\usepackage[normalem]{ulem}
\usepackage{soul}
\usepackage{colortbl,xcolor}
\usepackage{longtable,tabularx}

\hypersetup{
    colorlinks=true,
    linkcolor=blue,
    citecolor=blue,
    filecolor=magenta,      
    urlcolor=cyan,
    hypertexnames=true
    }

\usepackage{booktabs}       % professional-quality tables
\usepackage{nicefrac}       % compact symbols for 1/2, etc.
\usepackage{microtype}      % microtypography
\usepackage{ulem}

\title{Effects of Thom disk on alleviating ground effects of a wall-mounted rotating cylinder}

\author{Bao-Yuan Zhao$^1$\thanks{Corresponding author:kai.zhang@sjtu.edu.cn},
        Kai Zhang$^1$,
       	Dai Zhou$^1$,
        Shiliang Hu$^2$,
        Hanfeng Wang $^3$ \\
        $^1$School of Ocean and Civil Engineering, Shanghai Jiao Tong University, Shanghai 200240, China \\
        $^2$China Ship Scientific Research Center, Shanghai 200245, China \\
        $^3$School of Civil Engineering, Central South University, Changsha 410083, China
        }

\begin{document}

\maketitle
\section*{Nomenclature}

{\renewcommand\arraystretch{1.0}
	\noindent
	\begin{longtable*}{@{}l @{\quad=\quad} l@{}}
		
		$AR$ & aspect ratio \\
		$C_L,C_D$ & lift and drag coefficients \\	
		$D$ &    diameter of the cylinder\\
		$D_d$ &  diameter of the Thom disks\\
		$H$   &  height of the cylinder\\
		$P$   &  height of the secondary Thom disk\\
		$Re$  &  Reynolds number  \\
		$U_{\infty}$ & free-stream velocity\\
		$\alpha$ & velocity ratio \\	
		$\Omega$  & angular velocity \\
		
\end{longtable*}}

\section{Introduction}
Rotor sails, also known as Flettner rotors, were invented by Anton Flettner in 1925 \citep{flettner1925flettner} as a means of harnessing wind energy for ship propulsion. Recently, these devices have garnered renewed attention as a clean energy solution aimed at reducing emissions in the maritime industry.
A conventional Flettner rotor comprises an engine-driven, rotating cylinder mounted on a ship's deck, utilizing the Magnus effect to generate a lift force perpendicular to the flow direction \citep{seifert_review_2012}. This physical phenomenon has been the subject of extensive scientific investigation over the years.
Particular attention has been given to the aerodynamic performance of rotating cylinders, with studies exploring the influence of key parameters such as aspect ratios, velocity ratios, and Reynolds numbers \citep{rao2015review,shehata_wake_nodate,bordogna_experiments_2019,Cheng_Pullin_Samtaney_2018,CHEN2023115006}. These investigations have provided valuable insights into optimizing the performance and practical applications of rotor sails in maritime operations.

It is well acknowledged that the boundary conditions at the free ends of lifting surfaces induce significant three-dimensionality to the wake due to the formation of tip vortices \citep{zhang2020JFM, Pandi2023JFM}. These vortices arise from the pressure difference between the upper and lower surfaces of the lifting surface, generating a spiraling flow at the free end. This phenomenon leads to a substantial loss in lift and an increase in drag, thereby reducing overall efficiency.
For Flettner rotors, an effective method to mitigate these end effects is the addition of a Thom disk at the free end of the cylinder \citep{thom1934effect,liu2024pof}. The disk, mounted coaxially with the rotating cylinder, suppresses the free end effects by reorganizing the airflow around the cylinder's tip, ultimately enhancing the lift-to-drag ratio.
Further investigations by \citet{thouault_numerical_2012} and \citet{ritz_numerical_2015} have explored the aerodynamics of rotating cylinders equipped with multiple disks along the spanwise direction. 
Their findings revealed that these disks reduce the strength of the tip vortices, increase the streamwise velocity between the disks, and generate new vortices around the disks. Together, these effects lead to a marked improvement in the aerodynamic performance of the cylinder.

The ground effects pose another source of three-dimensionality to the flows over Flettner rotors. 
In the case of a stationary wall-mounted cylinder, numerous studies have revealed the complex vortex dynamics at the base \citep{cao2022JFM,liu2024pof}. A crucial feature is the development of a horseshoe vortex, which features two counter-rotating ``leg" vortices that wrap around the cylinder, and a ``head" vortex that extends upstream.
The inhomogeneity incurred by the ground also leads to oblique vortex shedding in regions away from the wall \citep{Williamson1989579,wang2006PoF,zhang2023OE}.
Regarding wall-mounted rotating cylinders, \citet{mittal_three-dimensional_2004} observed that at a low Reynolds number of 200, the wall effects induce the centrifugal instabilities along the span in an otherwise steady wake.
Recently, \citet{massaro_direct_2024} carried out direct numerical simulations of flows over Flettner rotors at $Re=3000$, revealing that the boundary layer interactions generate large-scale streamwise vortical structures that persist up to 60 diameters downstream.
Both studies have shown that the no-slip wall effects lead to an increase in drag and loss in lift, which adversely affects the aerodynamic efficiency of the Flettner rotors.

Building on the successful application of Thom disks in mitigating three-dimensional free-end effects, this study explores their potential to reduce adverse ground effects in Flettner rotors. 
To investigate this, we conduct three-dimensional direct numerical simulations of flows over wall-mounted rotating cylinders equipped with Thom disks at both the free end and near the ground. 
We mainly focus on the effects of the vertical location of the secondary Thom disk on the wake dynamics and aerodynamic forces of the rotating cylinder.
The insights obtained from this study can potentially inspire more efficient designs for Flettner rotors.

\section{Computational Setup}
\label{setup}

\begin{figure}
	\centering
	\includegraphics[width=.7\textwidth]{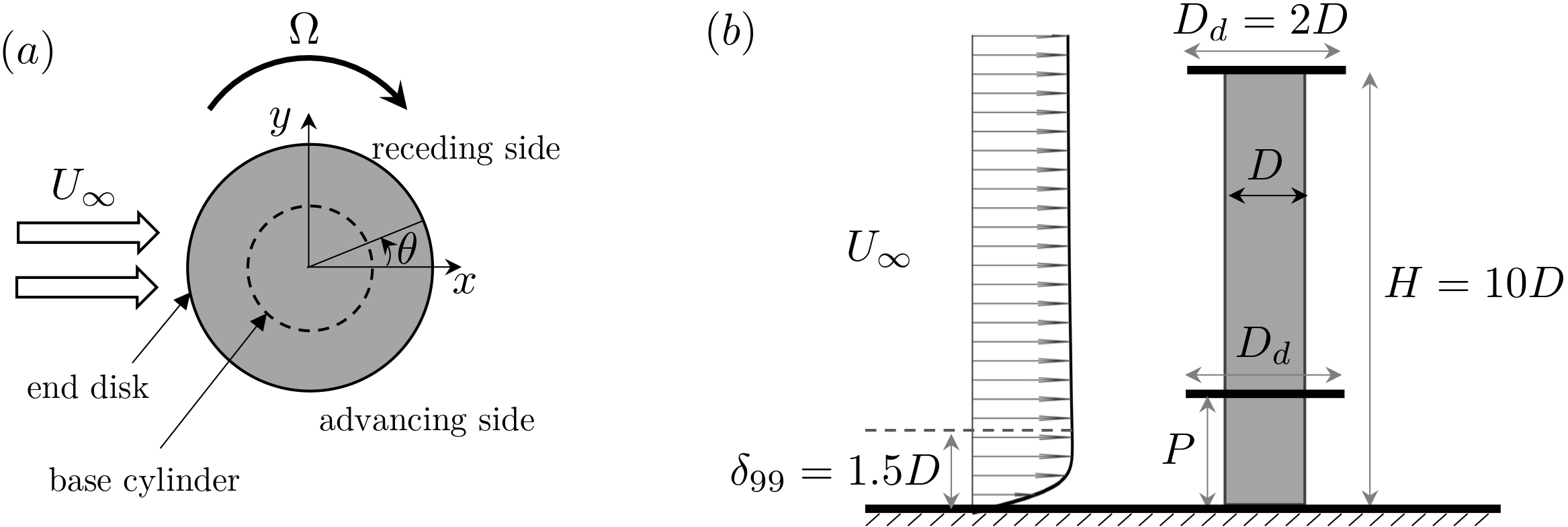}
	\caption{Sketch of the problem description. ($a$) top view and  (b) side view. The sketch is not to scale.}
	\label{fig:1}
\end{figure}

We consider three-dimensional incompressible flows over a Flettner rotor equipped with Thom disks, as shown in Fig. \ref{fig:1}. 
The $x$, $y$, and $z$-axes represent the streamwise, crossflow, and vertical directions, respectively. 
The rotating cylinder is characterized by a diameter of $D$ and a height of $H=10D$. 
The diameters of both the Thom disks are fixed at $D_d=2D$, which has been shown to effectively alleviate the free-end boundary effects \citep{thom1934effect,badalamenti2008effects,thouault_numerical_2012}.
The Thom disks are created with zero thickness to avoid meshing the side surfaces of the disks.
While the Thom disk at the free end is fixed, the location of the secondary one is parametrized by $P$, representing its distance to the ground. Thus, $P=0$ means the secondary disk is installed on the ground, and $P=10D$ suggests that the two Thom disks collapse into one at the free end.
The Reynolds number is defined based on the cylinder diameter $Re=U_\infty D/\nu$ (where $U_\infty$ is the freestream velocity, and $\nu$ is the kinematic viscosity of the fluid) and is fixed at 150.
The angular velocity of the rotating cylinder is set to be $\Omega=8$, which corresponds to a spin ratio of $\alpha=\Omega R/U_{\infty}=4$. Under this spin ratio and Reynolds number, the K\'arm\'an vortex shedding is suppressed even in the two-dimensional simulation, and the wake exhibits the Mode E instability as described in \cite{rao2015review}.
In what follows, the spatial variables are reported in their nondimensional forms by normalizing them with $D$, velocity by $U_\infty$ and time by $D/U_\infty$.

A rectangular computational domain covering $[x,y,z]\in[-30, 60]\times [-30,30]\times[0,40]$ is used in this study.
The cylinder, along with the Thom disks, and the bottom wall, are treated as no-slip boundaries, while the slip condition is applied to the side boundaries.  
A uniform velocity $U_{\infty}$ is set on the inlet boundary. 
Due to the no-slip boundary condition on the ground, this uniform inlet velocity progressively develops into a boundary layer as it propagates downstream.
At the location where the rotating cylinder is mounted, the boundary layer height is approximately $\delta_{99}=1.5D$.
The zero gradient condition is applied to the outlet, with a constant reference pressure $p=0$.
The lift and drag coefficients are defined as $C_{L,D}$=$F_{L,D}$/($\rho {U_{\infty}}^2 D H$/2), where $F_{L,D}$ are the lift and drag force acting on the cylinder, and $\rho$ denotes the fluid density.

\begin{figure}
	\centering
	\includegraphics[width=0.8\textwidth]{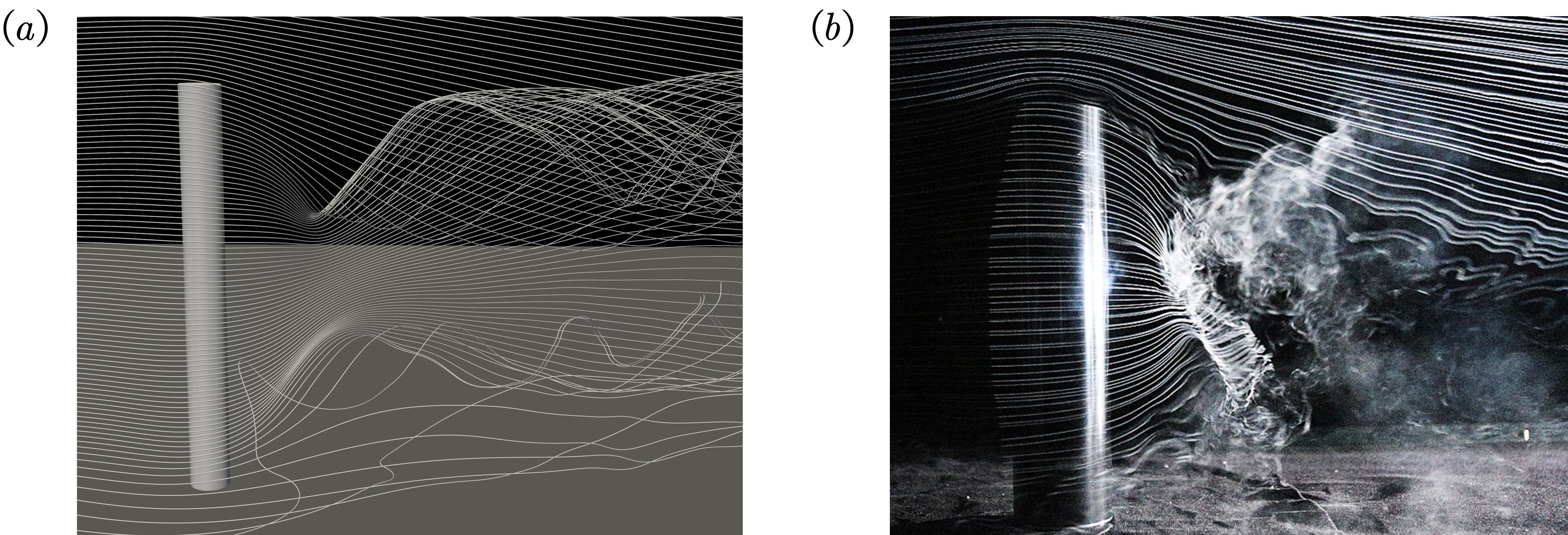}
	\caption{Streamline visualization of flow around the rotating cylinder in $(a)$ simulation and ($b$) wind tunnel experiment. In the numerical simulation, the wall-mounted cylinder rotates with $\alpha=4$ at $Re=150$. In the smoke visualization experiment conducted at the low speed wind tunnel of Central South University, the cylinder rotates with $\alpha=3.5$ at $Re=8\times 10^3.$ }
	\label{fig:2}
\end{figure}

Direct numerical simulations are performed using the unsteady  flow solver \emph{pimpleFoam} (\emph{OpenFOAM} package). 
It solves incompressible Navier-Stokes equations with second-order schemes in both space and time using the finite volume method.
The computational domain is discretized by hexahedron grids, with the mesh refined in the vicinity of the cylinder and its wake. The details of the mesh design are shown in table \ref{tab:meshtest}, along with a comparison with a refined mesh. 
It is observed that despite significantly increasing the mesh resolution in all three directions, the key aerodynamic forces are barely affected. Thus, the medium mesh is used for the current study.
We also compare the numerical results with the smoke visualization from a low-speed wind tunnel in Fig. \ref{fig:2}. 
Despite the large difference in the Reynolds number, our simulation is able to capture the major flow physics such as downwash from the free end and the upwash from the ground.

\begin{table}
	\centering
	\begin{tabular}{cccccccc}
		\hline
		case & $\Delta_r$ & $\Delta_{z}$ & $N_\theta$ & No. of grids & $\Delta t$ & $\overline{C_L}$ & $\overline{C_D}$ \\
		\hline
		medium & 0.006 & 0.003 & 240 & $3.2\times 10^7$ & 0.0025 & 8.732 & 5.031 \\
		refined & 0.0045 & 0.002 & 300 & $6.0\times 10^7$ & 0.001 & 8.750 & 5.025 \\
		\hline
	\end{tabular}
	\caption{Mesh dependency test for the flow over a wall-mounted rotating cylinder without Thom disks. $\Delta_r$ is the first layer grid size on the cylinder surface, $\Delta_z$ is the first layer grid size on the ground and the free end of the surface, as well as the Thom disk. $N_\theta$ is the number of grids along the circumferential direction. }
	\label{tab:meshtest}
\end{table}

\section{Results}
\label{sec:results}

\begin{figure}
	\centering
	\includegraphics[width=1.0\textwidth]{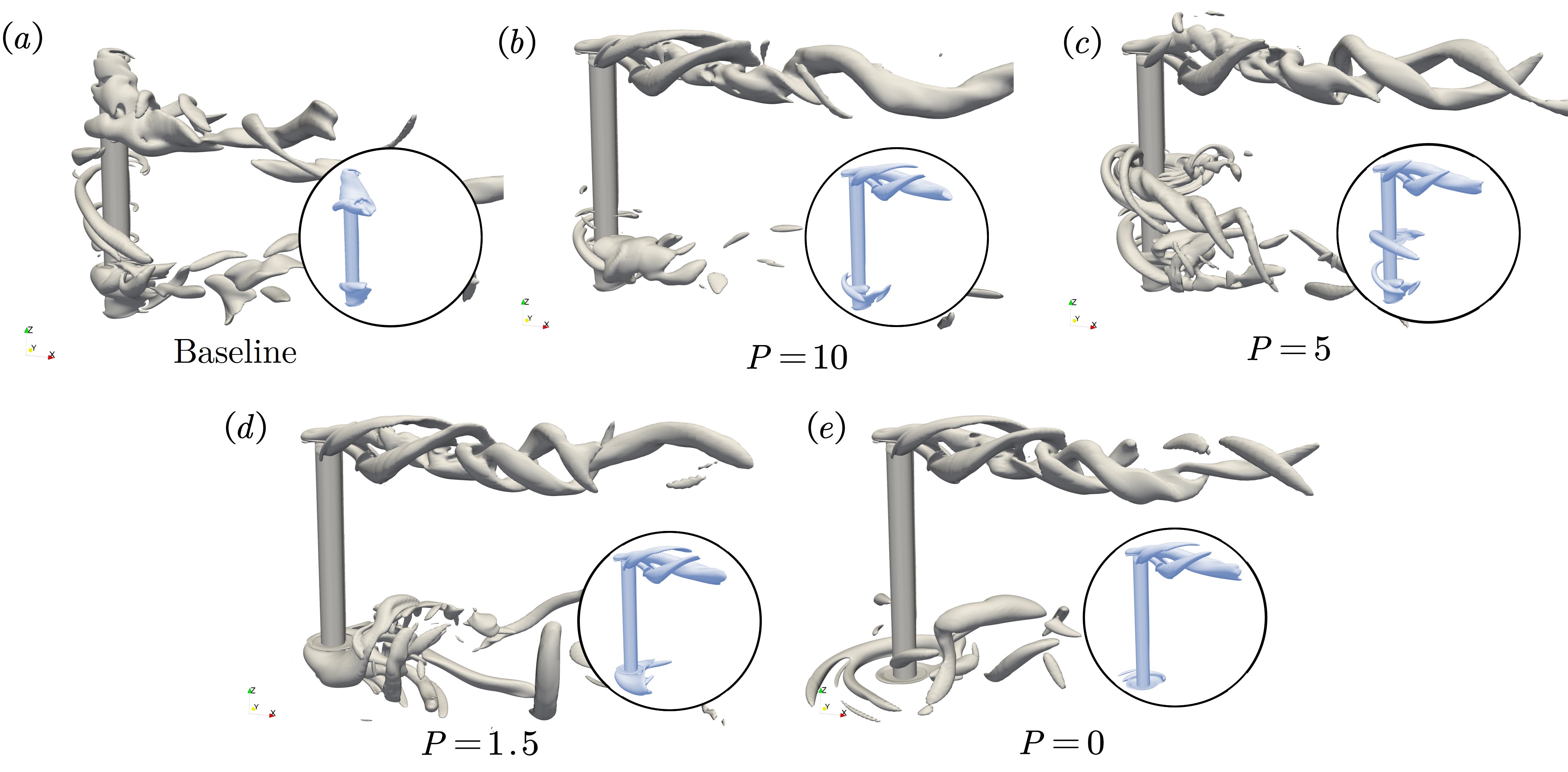}
	\caption{Vortical structures represented by an isosurface of $Q=1$ for different P. 
		The gray represents the instantaneous flow at $t=100$, and the blue represents the time-averaged flow.}
	\label{fig:3}
\end{figure}

\begin{figure}
	\centering
	\includegraphics[width=1\textwidth]{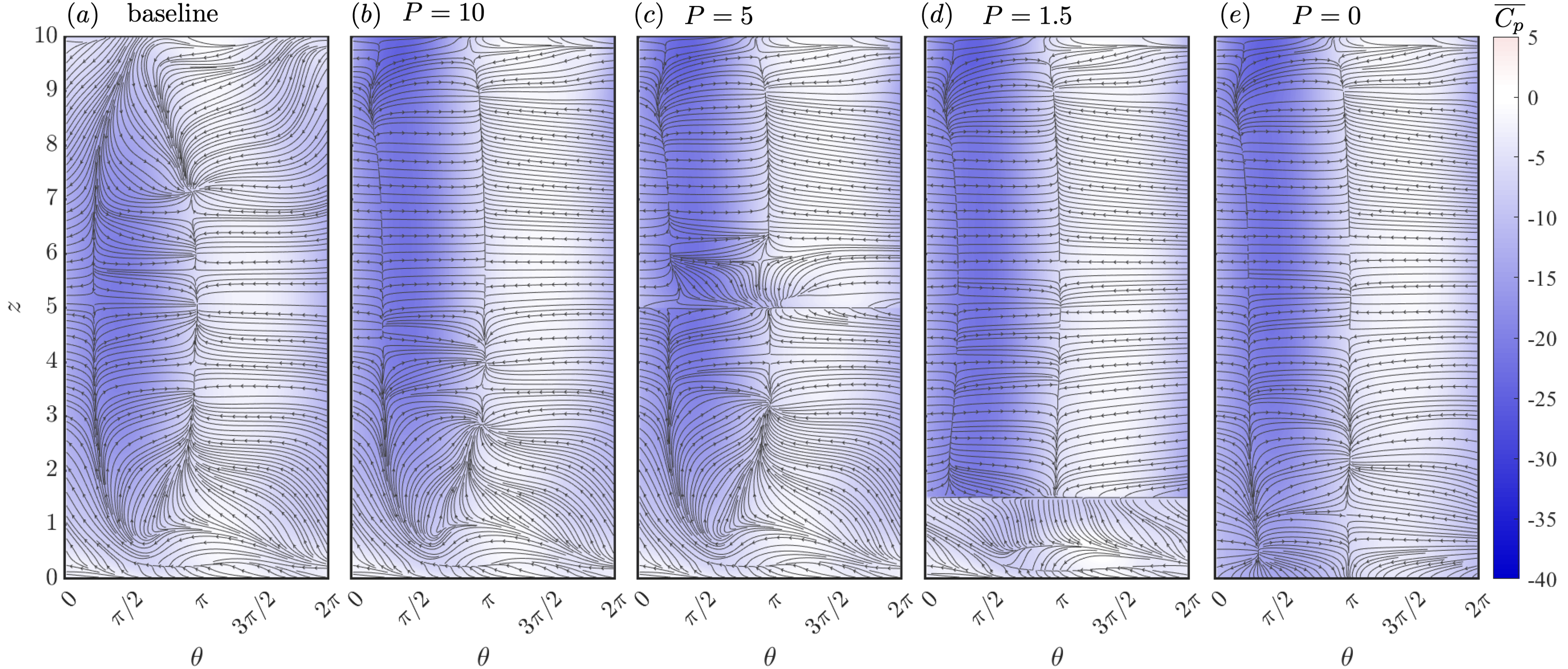}
	\caption{Skin friction line pattern for different $P$, colored by pressure coefficient on the cylinder surface.}
	\label{fig:4}
\end{figure}

We begin the discussion by examining the wake dynamics of the baseline case, i.e., without Thom disks. As shown in Figure \ref{fig:3}($a$), the wake structure of the wall-mounted rotating cylinder primarily consists of an ensemble of tip vortices near the free end and a base vortex system near the ground. Unlike the tip vortices typically observed in finite wings \citep{zhang2020JFM}, those generated by the free end of a rotating cylinder exhibit significantly more complex structures. A key distinction is the presence of multiple vortex tubes that helically interweave with each other as they are convected downstream. Furthermore, the tip vortices are deflected inward toward the cylinder due to the pronounced downwash effects induced by the free end. Near the ground, the base vortices exhibit a similarly complex topology. The conventional symmetric horseshoe vortex system observed in a stationary cylinder deforms into one-sided vortical structures due to the cylinder rotation. Driven by the upwash effect from the ground, these structures also lean towards the inward of the cylinder span. 
Additionally, the middle section of the cylinder surface is wrapped with small-scale Taylor-like vortices, which are typically observed in Taylor-Couette flows \citep{mittal_three-dimensional_2004,taylor1923PTRS}.

Fig. \ref{fig:4} shows the skin-friction line pattern ($\tau=\nu\boldsymbol{\omega}\times\boldsymbol{n}$, where $\boldsymbol{\omega}=\nabla\boldsymbol{u}$ is the vorticity vector, and $\boldsymbol{n}$ is the unit normal vector pointing outward from the cylinder surface) on the cylinder surface, together with the distribution of pressure coefficients $C_p=(p-p_\infty)/(0.5\rho U_\infty^2)$.
In general, the region around $\theta\approx \pi/2$ (i.e., the receding side) features accelerated boundary layer flow, as indicated by the streamwise skin friction shown in Fig. \ref{fig:4}. 
This is also the region where high negative pressure resides, providing a strong suction force that generates lift.
In the baseline case, both the free end and the ground induce three-dimensionality to skin-friction line patterns. 
The negative pressure region shrinks significantly near the boundaries.

With the addition of the Thom disk at the free end (i.e., $P=10$), the tip vortical structures appear stronger, due to the increased circular velocity at the periphery of the disk, as shown in Fig. \ref{fig:3}(b).
In contrast to the baseline case, these tip vortical structures suffer less from the downwash effects.
The skin friction line for this case exhibits increased uniformity near the free end, as shown in Fig. \ref{fig:4}($b$), and the negative pressure extends all the way up to the free end.
As a result of the two-dimensionalization effect from the end Thom disk, the Taylor-like vortices that are prevalent in the baseline case are also suppressed.

Installing a secondary Thom disk at half height ($P=5$) of the rotating cylinder introduces strong vortex shedding near the disk. 
These vortical structures reside mainly on the advancing side of the disk, forming interwoven vortices that are reminiscent of short-wavelength instability observed in rotating disks at incidence \citep{lee2022instabilities}.
On the cylinder surface, this secondary Thom disk introduces additional three-dimensionality to the flow, as shown in Fig. \ref{fig:4}($c$).
As the height of the secondary Thom disk is decreased to $P=1.5$, the flow structures generated by the disk and the ground amalgamate, forming a large-scale base vortex system.
Fig. \ref{fig:4}($c$) reveals that the region below the secondary Thom disk exhibits significant three-dimensional flow pattern, while the flow in the rest of the span is two-dimensionalized.
When the secondary Thom disk is installed on the ground, the rotating disk generates clear horseshoe-like vortices at its windward part.
As shown in Fig. \ref{fig:4}($e$), the three-dimensionality is significantly reduced on most parts of the cylinder surface, leading to a more uniform distribution of the negative pressure along the cylinder span. 

\begin{figure}
	\centering
	\includegraphics[width=0.9\textwidth]{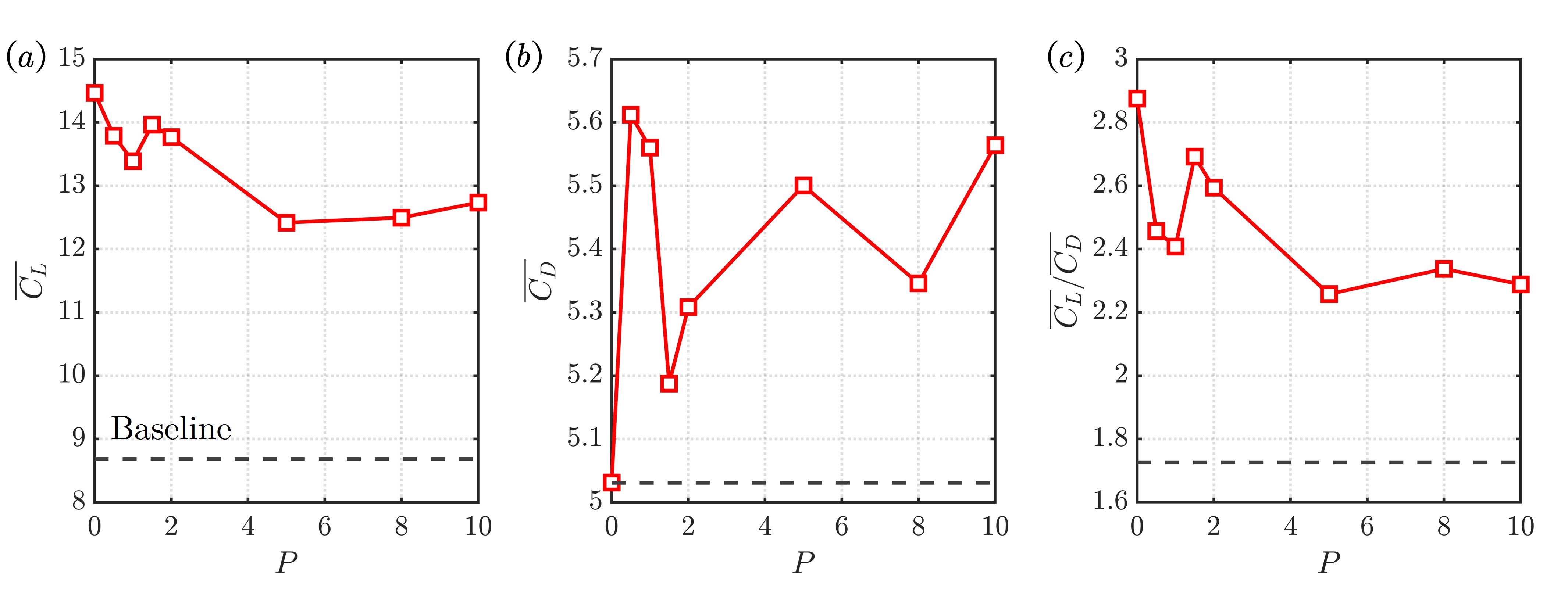}
	\caption{Time-averaged $C_L,C_D$ and lift-to-drag ratio at various $P$, the dotted line represents when there are no disks}
	\label{fig:5}
\end{figure}

The aerodynamic force coefficients of the cylinder-disk bundle as a function of the height of the secondary Thom disk are presented in Fig. \ref{fig:5}. 
The baseline case suffers from strong end effects from both the free side and ground side, and features a reduced lift of $\overline{C_L}=8.75$.
The addition of the end Thom disk significantly improves the lift coefficient by around 50\%, at the cost of slightly increased drag.
A considerable increase in the lift is achieved when the secondary disk is installed on the ground, due to the strong two-dimensionalization of the surface flow as discussed in Fig. \ref{fig:4}($e$). 
Another benefit of this ground Thom disk is that it offsets the increase in drag due to the end disk, resulting in almost the same drag as the baseline case.

As the height of the secondary disk increases, $\overline{C_L}$ declines in general, accompanied by a secondary peak around $P=1.5$, which is roughly the height of the boundary layer in this case.
When $P\gtrsim 5$, the secondary Thom disk presents negative effects on lift generation, due to the additional three-dimensionality introduced by it.
In general, compared to a single Thom disk at the free end, installing a secondary Thom disk near the ground improves the lift-to-drag ratio of the Flettner rotor. 
This observation lays the foundation for designing more efficient WASP system that aid the decarbonization of modern ships.

\section{Conclusions}
\label{sec:conclusion}
In this study, we have investigated the use of Thom disks to reduce the adverse ground effects on Flettner rotors through direct numerical simulations. We found that the addition of a secondary Thom disk significantly alleviates the three-dimensional flow structures induced by the ground, improving the aerodynamic performance of the rotor. Placing the disk directly on the ground optimizes the lift-to-drag ratio by creating a more uniform pressure distribution along the rotor’s surface. 
Our results provide valuable insights for optimizing the design of Flettner rotors, offering a pathway to more efficient and sustainable maritime propulsion systems. Future work will focus on exploring the performance of the Thom disks under a wider range of operational conditions.

\section*{Acknowlegement}
The financial support from National Natural Science Foundation of China (No. 12202271) is gratefully acknowledged.

\bibliographystyle{unsrtnat}
\bibliography{reference}

\end{document}